# Spin State Disproportionation in Insulating Ferromagnetic LaCoO$_3$ Epitaxial Thin Films

Shanquan Chen,[1] Jhong-Yi Chang,[2] Qinghua Zhang,[3] Qiuyue Li,[4,5] Ting Lin,[3] Fanqi Meng,[3] Haoliang Huang,[6] Shengwei Zeng,[7] Xinmao Yin,[8] My Ngoc Duong,[2] Yalin Lu,[6] Lang Chen,[9] Er-Jia Guo,[3] Hanghui Chen,[5,10] Chun-Fu Chang,[11] Chang-Yang Kuo,[2,12,*] Zuhuang Chen[1,13,*]

[1]*School of Materials Science and Engineering, Harbin Institute of Technology, Shenzhen, 518055, China*
[2]*Department of Electrophysics, National Yang Ming Chiao Tung University, Hsinchu, 30010, Taiwan*
[3]*Beijing National Laboratory for Condensed Matter Physics and Institute of Physics, Chinese Academy of Sciences, Beijing 100190, China*
[4]*Department of Electronic Science, East China Normal University, 200241, Shanghai, China*
[5]*NYU-ECNU Institute of Physics, NYU Shanghai, Shanghai 200122, China*
[6]*Hefei National Research Center for Physical Sciences at the Microscale and Anhui Laboratory of Advanced Photon Science and Technology, University of Science and Technology of China, Hefei 230026, China*
[7]*Department of Physics, Faculty of Science, National University of Singapore, Singapore 117551, Singapore*
[8]*Shanghai Key Laboratory of High Temperature Superconductors, Physics Department, Shanghai University, Shanghai 200444, China*
[9]*Department of Physics, Southern University of Science and Technology, Shenzhen, 518055, China*
[10]*Department of Physics, New York University, New York, NY, 10012, USA*
[11]*Max-Planck Institute for Chemical Physics of Solids, Nöthnitzer Str. 40, 01187 Dresden, Germany*
[12]*National Synchrotron Radiation Research Center, 101 Hsin-Ann Road, Hsinchu, 30076, Taiwan*
[13]*Flexible Printed Electronics Technology Center, Harbin Institute of Technology, Shenzhen, 518055, China*

[*]Corresponding authors: zuhuang@hit.edu.cn; changyangkuo@nycu.edu.tw

**Abstract**

The origin of insulating ferromagnetism in epitaxial LaCoO$_3$ films under tensile strain remains elusive despite extensive research efforts have been devoted. Surprisingly, the spin state of its Co ions, the main parameter of its ferromagnetism, is still to be determined. Here, we have systematically investigated the spin state in epitaxial LaCoO$_3$ thin films to clarify the mechanism of strain induced ferromagnetism using element-specific x-ray absorption spectroscopy and dichroism. Combining with the configuration interaction cluster calculations, we unambiguously demonstrate that Co$^{3+}$ in LaCoO$_3$ films under compressive strain (on LaAlO$_3$ substrate) are practically a low spin state, whereas Co$^{3+}$ in LaCoO$_3$ films under tensile strain (on SrTiO$_3$ substrate) have mixed high spin and low spin states with a ratio close to 1:3. From the identification of this spin state ratio, we infer that the dark strips observed by high-resolution scanning transmission electron microscopy indicate the position of Co$^{3+}$ high spin state, *i.e.*, an observation of a spin state disproportionation in tensile-strained LaCoO$_3$ films. This consequently explains the nature of ferromagnetism in LaCoO$_3$ films.

## I. INTRODUCTION

Perovskite oxides exhibit a variety of electronic properties, including high-temperature superconductivity, colossal magnetoresistance, multiferroism, as a consequence of complex interplay among lattice, charge, spin, and orbital degrees of freedom [1-6]. Epitaxial strain can dramatically alter the physical properties of the strongly correlated oxide materials [7]. One notable example is the tensile-strain-induced ferromagnetism in LaCoO$_3$ (LCO) epitaxial thin films [8-10]. Bulk LCO is a diamagnetic insulator below ~100 K due to a low spin (LS, $t_{2g}^6 e_g^0$, S=0) ground-state configuration of Co$^{3+}$, and becomes paramagnetic at higher temperature with increasing population of Co$^{3+}$ high spin (HS, $t_{2g}^4 e_g^2$, S=2) state [11]. Although this intriguing system has been studied extensively for decades [9,12-17], the origin of insulating ferromagnetism in tensile-strained LCO films still remains controversial. Previous investigations have proposed several competing models for the mechanism governing the ferromagnetism in the LCO films, including orbital order composed of Co$^{3+}$ intermediate spin (IS, $t_{2g}^5 e_g^1$, S=1 [18]) and/or Co$^{3+}$ HS state [13,19], superexchange-like interaction between LS Co$^{3+}$ and HS Co$^{2+}$ induced by oxygen vacancies [14,15], spin-state order among Co$^{3+}$ LS and HS [12,20] or IS state [21], charge order/disproportionation [22], and intrinsic twin domains with HS/IS and LS state order [23,24]. Apparently, the spin state of Co ions in strained LCO films is a key factor for its ferromagnetism. Yet, its spin states are controversially reported.

Here, to understand the nature of insulating ferromagnetism in epitaxial LCO films, we have engaged a systematic study using soft x-ray absorption spectroscopy (XAS), x-ray linear dichroism (XLD) spectroscopy, magnetic circular dichroism (XMCD) spectroscopy and high-resolution scanning transmission electron microscopy (STEM). Combining with the configuration cluster-interaction (CI) calculations, we reveal that the $Co^{3+}$ ions in films under tensile strain (on $SrTiO_3$, STO) consist of both the LS and HS states, and, importantly the ratio of HS to LS in cobalt ions is about 1:3, whereas the $Co^{3+}$ ions in films under compressive strain (on $LaAlO_3$, LAO) possess mainly a LS state. This HS to LS ratio of 1:3 naturally explains the low net magnetic moment of 1 $\mu_B$/Co in the tensile strained films. Microscopically, we investigated the origin of the commonly observed stripe patterns in STEM images with different electron dose conditions. With a high electron dose, we observed a dark- to bright-strip ratio of ~1:3 in the tensile strained LCO/STO films, whereas with a low electron dose a homogeneously bright image was observed. The images from the LCO/LAO films are electron dose independent and have no dark stripes. This indicates that the dark stripes could be the area of HS state, but the formation of the dark stripes due to oxygen vacancies ordering is extrinsic. Important is that there is an intrinsic HS- and LS-state disproportionation in the LCO/STO films. This $Co^{3+}$ HS and LS disproportionation provides a particular ferromagnetic coupling of the $Co^{3+}$ HS intermediated by the $Co^{3+}$ LS for the ferromagnetism in the tensile strained films.

## II. EXPERIMENTAL RESULTS

High-quality LCO films with a thickness of 12 nm were grown on [001]-oriented STO and LAO substrates by pulsed laser deposition (see details in Sec. IV). All LCO films have smooth surfaces and coherently strained to the underlying substrates (Fig. 1 and see Supplemental Material Fig. S2[18]). Only (00*l*) diffraction peaks from the films and substrates are visible, which is indicative of epitaxial growth without impurity [Fig. 1(a)]. Combined with the results of the x-ray reciprocal-space mappings (RSM) along the pseudocubic (013) direction [Fig. 1(b-c)], the out-of-plane lattice parameter of the LCO films shrink (expand) when the films are epitaxially grown on tensile-STO (compressive-LAO) substrates because of the Poisson effect. Fig. 2(a) displays the SQUID magnetometry results for the films. Similar to bulk LCO, the film grown on LAO does not show any discernible ferromagnetic transition or hysteresis loop [Fig. 2(a)]. In contrast, LCO grown on STO shows a distinct magnetic transition at 80 K and a clear ferromagnetic hysteresis loop with saturation magnetization of ~1 $\mu_B$/Co at 10 K, indicates a ferromagnetic order with $T_C \sim$ 80 K. Furthermore, the transport measurement reveals that the LaCoO$_3$ films show insulating behavior [Fig. 2(b)]. The macroscopic magnetic and transport measurement results are consistent with previous studies [9,15,23,25-27].

  To figure out the spin states in LCO films, we measured the XAS at Co $L_{2,3}$ edges. The XAS at the *L* edges measure the excitations of electrons from the 2*p* core levels to the 3*d* unoccupied states. The strong Coulomb interaction between a 2*p* core-hole and 3*d* electrons narrows the 3*d* bandwidth and enhances the local solid-state effect. It makes the XAS a suitable tool to investigate the local spin state and orbital occupation [10,28-30]. Fig. 3 shows the experimental Co $L_{2,3}$ XAS spectra taken from LCO films grown on STO and LAO substrates at 30 K. Obviously, Co$^{3+}$ is the main valence state of Co ion in both films. Besides, clear spectral differences between the XAS spectrum of LCO/STO and LCO/LAO are observed (inset of Fig. 3). It is important to note that the XAS difference shown in Fig. 3 for LCO between the two different strain states is similar to previous results on bulk LCO measured at high and low temperatures, respectively [10]. It indicates a spin-state difference of Co ions for LCO films under compressive strain and tensile strain.

To look for insight into XAS, we simulate the XAS spectra of $Co^{3+}$ ions with different spin states by employing the CI cluster calculations with full atomic multiplet theory. We first calculated the XAS spectra for $Co^{3+}$ ions in pure LS and pure HS states (Fig. 3). One can see that the calculated XAS spectra for pure HS and LS states nicely reproduce the experimental XAS of $Sr_2CoO_3Cl$ and $EuCoO_3$ (Fig. 3), in which the former compound has $Co^{3+}$ ions in a pure HS state but the later in a pure LS state [28-30]. We next did the linear combination of HS and LS XAS spectra to fit the experimental XAS spectra of LCO films. The curves in the upper-right inset of Fig. 3 show the experimental XAS spectra and theory spectra calculated with a pure LS state (blue) and with HS: LS~1:3 mixed spin states (red). One can see that the theoretical spectra are in good agreement with the experimental ones. The results indicate that $Co^{3+}$ in LCO/LAO contains mainly a LS state, while $Co^{3+}$ in LCO/STO is a mixed spin state with a HS to LS ratio of 1:3.

To further confirm the $Co^{3+}$ spin state in the LCO films, we measured the XLD spectra in grazing incidence geometry[Fig. 4(a)]. In principle, the XLD originates from two contributions, namely the orbital anisotropy and the magnetic contribution. The XLD spectra shown in Fig. 4 were collected after LCO cooling down from room temperature to T = 30 K without applying a magnetic field. One can thus expect no long-range magnetic order presenting in LCO and the contribution from magnetism to XLD is minimal. That is, the orbital anisotropy should be the main origin for causing the XLD signal. Under the LS state, six electrons fully occupy six $t_{2g}$ orbitals, making the ground state wavefunction nearly spherical symmetry. There is thus no orbital anisotropy and the XLD is expected to be zero for $Co^{3+}$-ion under the LS state. In Fig. 4b, one sees that the XLD signal of $Co^{3+}$-ion in the compressive LCO film is negligibly small, indicating essentially no orbital anisotropy for the LS state. The XAS spectra calculated for $E//c$ and $E//ab$ shown in Fig. 4b under a pure LS state generates no XLD signal, which reproduces the experimental results and reveals the nearly 100% LS state for $Co^{3+}$-ion in the compressive LCO/LAO film.

In contrast, as shown in Fig. 4c, the tensile-strained LCO/STO film exhibits clear linear dichroism. For a HS state of $Co^{3+}$-ion, five majority spin electrons, half-filling the 3$d$ orbitals, make a spherical symmetry of wavefunction, and a minority spin electron occupying a specific orbital can cause an orbital anisotropy and give rise to XLD signals for the film. The XLD signal in Fig. 4c clearly shows that the intensity of XAS for $E//c$ is stronger than that of $E//ab$. This indicates that a portion of $Co^{3+}$ cations are a HS state in the tensile-strained LCO film, and the orbital occupied by the minority spin electron of the $Co^{3+}$ ions is in the $ab$-plane of the film. To address the HS contribution to XLD, we calculated the XAS spectra of HS $Co^{3+}$ ions by considering the energy of $d_{xy}$ orbital lower than that of $d_{xz}/d_{yz}$ by ~50 meV. A minority spin electron of HS $Co^{3+}$ ions thus occupies the $d_{xy}$ orbital and causes the orbital anisotropy, which fully accords to the tensile-strain status of the LCO/STO film. To simulate the experimental XLD spectrum of LCO/STO, we have carried out a simple simulation by making a superposition of the calculated HS and LS XLD spectra. We found that the spectra with a HS to LS ratio of 1:3 reproduce the experimental spectra the best as shown in Fig. 4c, which is consistent with the ratio obtained for simulating the isotropic XAS spectrum. Another piece of important information is that this occupied $d_{xy}$ orbital of HS $Co^{3+}$ may explain the in-plane magnetic easy axis in LCO/STO films [31].

The presence of $Co^{3+}$ HS in LCO/STO films offers the essential for its ferromagnetism. To clarify the contribution of $Co^{3+}$ HS state to ferromagnetism, we carried out remanent XMCD experiments on magnetized LCO/LAO and LCO/STO at T = 30 K [Fig. 4(d)]. Figs. 4e and f show experimental XAS spectra taken with circularly polarized light on the LCO films. The XMCD measurement demonstrates a consistent result with the above macroscopic magnetic properties measured in magnetometry. Spectra of LCO/LAO [Fig. 4(e)] shows an almost zero dichroism signal, indicating no spontaneous magnetization. Spectra of LCO/STO [Fig. 4(f)] exhibits a clear dichroism signal, addressing a net ferromagnetic order of the $Co^{3+}$ ions at 30 K. To address the origin of the XMCD, we again simulate the XMCD spectra by CI cluster calculations. We take the same ratio of two spin states found by both the XAS and XLD simulations, a pure LS state for LCO/LAO film and HS:LS=1:3 for LCO/STO film, to calculate the XMCD spectra. The results are also included in Fig. 4. It is not surprising to see that the calculated XMCD for LCO/LAO is zero because of a pure LS (S=0) state. The calculation for LCO/STO with HS:LS=1:3 exhibits a clear XMCD signal and the calculated spectra nicely reproduce the experimental spectra. Hence, it is the HS $Co^{3+}$-ions which contribute to the XMCD signal and form the ferromagnetic order. It is worthwhile to mention that this HS to LS ratio of 1:3 naturally explains the low net magnetic moment of ~1 $\mu_B$/Co in the LCO/STO films from the above SQUID magnetometry results. We also use first-principles calculations to study the LS state of LCO under LAO substrate and a mixed-spin state (HS:LS=1:3) of LCO under STO substrate. In both spin configurations, an insulating state is found in the calculations. In the mixed-spin state, calculations also find a net magnetization of ~1 $\mu_B$ per formula (see Supplemental Material Fig. S3[18]).

So far, using high resolution STEM lattice modulations in the form of atomic-scale stripe patterns were commonly observed [14,15,23,24,27,32-34]. However, not only the origin of the dark stripes observed in LCO/STO has been debated [14,23,24,27,32,33], but also various spin state models were proposed [14,24,32]. To clarify the relationship between the microstructure and the $Co^{3+}$ spin state, we performed high-resolution STEM measurements on LCO/LAO and LCO/STO films (Fig.5 and see Supplemental Material Fig. S4[18]). We paid especial attention to the used electron dose, since STEM technique is not a non-destructive technique, a high-energy electron beam could damage/reduce the LCO thin films to a certain extent, resulting in a change in its stoichiometry [35]. Fig. 5 shows the high-angle angular dark-field (HAADF) STEM images of different stages of the beam exposure experiment on LCO/STO. Under low-dose electron beam irradiation (~$10^6$ e-/angstrom$^2$), no dark stripes are seen in the STEM image (Fig. 3a). However, as the electron beam irradiation dose increases (~$10^8$

e-/angstrom$^2$), we observed clearly ordered dark stripes in the STEM image [Fig. 5(b)]. According to statistics, the ratio of the unit cells where the dark and bright stripes are located is about 1:3 [Fig. 5(c)]. In contrast, no dark stripes were observed in STEM images for compressive LCO film grown on LAO under both low-dose and high-dose electron beam irradiation conditions (same irradiation as LCO/STO) (see Supplemental Material Fig. S4[18]). These results are consistent with previous literature reports [14,15,23,24]. This invites a comparison with the HS-LS ratio of 1:3 found by the above mentioned XAS results. It is reasonable to speculate that the position of the dark stripes is not random, but rather the place where the high spin state $Co^{3+}$ is. We have further verified this assignment by using electron-energy-loss spectroscopy (EELS) (see Supplemental Material Fig. S5[18]). For LCO/LAO, there is no difference in the shape, intensity and peak position of the O-$K$ and Co-$L$ edges of bright stripes in different regions. However, the clear differences between the EELS in the dark (black line) and bright stripes (red line) in the LCO/STO film are seen [18]. The front peak (light golden area) intensity of the O-$K$ edge corresponding to the dark stripes is lower than that of the bright stripes pattern, which indicating the formation of oxygen vacancies along these places. In addition, the dark stripes have a higher $L_3/L_2$ peak intensity ratio than bright stripes at the Co-$L$ edges [18]. These findings indicate that the oxidation state of Co along the dark stripes has been reduced by a relatively high dose of electron beam. It is well-known that the metal ion-ligand bond of HS is weaker than that of LS [36,37], therefore, the oxygen ions around the HS are likely to be knocked out more easily. This HS-dark stripe and LS-bright stripe assignment is also supported by the bond-length modulation observation by STEM studies as the ionic radius of HS $Co^{3+}$ is larger than that of LS $Co^{3+}$ [32]. We conclude that the commonly observed dark stripes in LCO/STO films is an extrinsic reduction of oxygen by a high dose electron beam used in STEM. Yet, most importantly, it gives an access of a direct atomic-scale positioning of HS $Co^{3+}$ ions, evidencing an intrinsic HS and LS disproportionation in LCO/STO films.

We in fact restore the HS and LS disproportionated insulating phase for bulk LCO at intermediate temperature [38]. The advance of LCO/STO film is that by strain engineering, this HS and LS disproportionation is realized at low temperature or as the ground state. The spin state of the $Co^{3+}$ ions in LCO is mainly determined by the competition between the crystal-field splitting $\Delta_{CF}$ and the Hund's exchange interaction $J_H$ [10]. The LCO film under the tensile strain has an average Co-O bond length longer than that of one under compressive strain [39]. The average Co-O bond length determines the magnitude of $\Delta_{CF}$: a longer Co-O bond length would cause a weaker $\Delta_{CF}$. Apparently, the tensile strain enforced by STO is on the board of LS and HS formation, so the $Co^{3+}$ ion can be either LS or HS[20].

The HS and LS state disproportion in LCO/STO plays an important role on its magnetism [12,20]. Fig. 6 displays a sketch of the superexchange-like process of $t_{2g}$ and $e_g$ electrons in a HS-LS-HS configuration for both ferromagnetic and antiferromagnetic couplings. For the ferromagnetic coupling, the final state of the virtual hopping process ends up in the same spin state configuration of the initial state [Fig. 6(a)]. Whereas the antiferromagnetic coupling brings the system on a different spin state configuration [Fig. 6(b)]. The ferromagnetic coupling is energetically more favorable than the antiferromagnetic coupling by $8J_H$. We note that this energetic superiority of the ferromagnetic coupling of $Co^{3+}$ ions stand also for configurations such as HS-LS-LS-HS and HS-LS-LS-LS-HS and so on, once the $Co^{3+}$ HS is intermediated by the LS.

## III. SUMMARY

In summary, using a combination of element-specific x-ray absorption spectroscopy and dichroism together with the configuration interaction cluster calculations, we unambiguously demonstrate that Co ions in LCO/LAO thin films under compressive strain are practically a low spin state, whereas Co ions in LCO/STO thin films under tensile strain have mixed high spin and low spin states with a ratio close to 1:3. From the identification of this spin state ratio, we infer that the dark strips in LCO/STO observed by STEM indicate the position of $Co^{3+}$ high spin state. Yet, these dark strips are extrinsically caused by the high dose of electron beam used in STEM. Most importantly, this provides an access to experimentally observe the $Co^{3+}$ HS and $Co^{3+}$ LS state disproportionation in LCO/STO. In such a HS and LS state disproportionation, the ferromagnetic coupling of the $Co^{3+}$ HS intermediated by $Co^{3+}$ LS is energetically more favorable than the antiferromagnetic coupling.

## IV. METHODS

### A. LaCoO$_3$(LCO) films growth and structural characterization

High quality LCO films (12~15 nm) were grown on [001]-oriented single crystal SrTiO$_3$(STO) and LaAlO$_3$(LAO) substrates by pulsed laser deposition (Arrayed Materials RP-B). During the film deposition, the KrF excimer laser ($\lambda = 248$ nm) energy density and ablation repetition were given 1.5 J/cm$^2$ and 3 Hz, respectively. The deposition temperature was kept at 700 °C, and the oxygen pressures during film deposition was maintained at 20 Pa. All films were cooled down (10 °C/min) to room temperature with 10000 Pa of oxygen atmosphere to ensure the oxygen stoichiometry. High-resolution x-ray diffraction (XRD) measurements were performed on a Rigaku-Smartlab, 9 KW diffractometer with Cu $K_{\alpha 1}$ radiation. The morphology was determined by an Asylum Research MFP-3D-Infinity atomic force microscopy (AFM).

**B. Electronic transport and Magnetic measurements**

The magnetic properties of the LCO films were measured using a Quantum Design Superconducting quantum interference device (SQUID) magnetometry measurements. The Magnetization-temperature ($M$–$T$) curves were measured at the magnetic field of 100 Oe applied along the films plane (in plane) after the field cooling in the same field. The in plane magnetization-magnetic field hysteresis loops ($M$-$H$ curves) was measured at 10 K with magnetic field up to ±6 T. DC transport measurements were carried out in a Quantum Design Physical Property Measurement System (PPMS) from 200 to 380 K under zero magnetic field.

**C. Soft X-ray spectroscopy measurements**

X-ray spectroscopy measurements were carried out in the total electron yield mode (TEY) at the TLS11A and TPS45A beamline of the National Synchrotron Radiation Research Center (NSRRC) in Taiwan. The X-ray absorption spectroscopy (XAS) measurements were performed at 30 K temperature without applied magnetic field using TEY mode and an X-ray angle of incidence of 20° to the sample surface. X-ray linear dichroism (XLD) measurements were performed at 30 K and were obtained from the difference of horizontal and vertical polarized light absorption spectra without applied magnetic field using TEY mode. The X-ray beam was incident on the sample at an angle of 20° from the sample surface. X-ray magnetic circular dichroism (XMCD) spectra were measured in the TEY mode without applied magnetic field using fixed X-ray circular polarization at 30 K in grazing incidence geometry. Before the XMCD measurement, the LCO films are first magnetized by a magnetic field of 1 Tesla.

**D. High-resolution STEM characterizations**

For cross-sectional microscopy, sample was prepared by using focused ion beam (FIB) milling. Cross-sectional lamellas were thinned down to 60 nm thick at an accelerating voltage of 30 kV with a decreasing current from the maximum 2.5 nA, followed by fine polish at an accelerating voltage of 2 kV with a small current of 40 pA. The atomic scale HAADF-STEM images of $LaCoO_3$ films were performed by Cs-corrected JEM ARM200CF microscope operated at 200 kV using a high-angle annular detector for Z-contrast imaging; and the beam convergence angle was 28.5 mrad and a collection angle of 90 – 370 mrad.

**E. Configurational cluster-interaction (CI) calculations**

The calculation employs the fully atomic multiplet effect and configurational interaction as well as local solid effect which includes the local crystal field, hybridization between Co and its ligand atoms, mean filed exchange interaction, and spin orbital coupling. Parameters (in eV) used in the spectra fitting lists as following: $\Delta=2.0$, $U_{dd}=5.5$, $U_{pd}=7.0$, $F2(dd)=12.662$, $F4(dd)=7.916$, $F2(dp)=7.899$, $G1(dp)=5.947$, $G3(dp)=3.384$, $SOCC(d)=0.074$, $SOCC(p)=9.746$, Slater Integral reduction = 0.75, 10Dq(for LS)=0.62, 10Dq(for HS)=0.2.

**F. First-principles calculations**

we perform first-principles calculations using the projector augmented wave method[40], as implemented in the Vienna Ab initio Simulation Package (VASP)[41]. The cutoff energy is 600 eV. The self-consistent convergence criterion is $10^{-6}$ eV. The force convergence criterion is 0.01 eV/ Å. The pressure convergence threshold on the simulation cell is 1 kbar. We use the generalized gradient approximation with the Perdew-Burke-Ernzerhof parameterization (GGA-PBE) as the exchange-correlation functional[42]. We use a 9 × 9 × 6 Monkhorst-Pack grid[43] as the Brillouin zone integration. To model the correlation effects on Co $3d$ orbitals, we use the rotationally invariant Hubbard $U$ method[44]. Following the previous study[45], we find that a $U_{Co}$ = 6 eV reproduces the insulating properties of LaCoO$_3$ thin films in both low-spin, high-spin and mixed-spin states. We also turn on $U_{La}$ = 11 eV for La-4$f$ orbitals so that they are moved away from the Fermi level. Bulk LaCoO$_3$ has a crystal structure of space group $R\bar{3}c$. To simulate epitaxial strain, we follow the previous study and adopt a tetragonal conventional cell using the starting ionic positions as mapped from bulk LaCoO$_3$. We define $a$ and $b$ as the in-plane lattice parameters, whereas $c$ is the out-of-plane lattice parameter. To simulate the epitaxial LaCoO$_3$ thin films on different substrates, we constrain the in-plane lattice constant $a=b=5.57$ Å (SrTiO$_3$) and $a=b=5.39$ Å (LaAlO$_3$), respectively. The out-of-plane lattice constant $c$ and all the internal coordinates are fully relaxed. The conventional cell has four formula units (20 atoms). In the low spin (LS) configuration, all the cobalt atoms are low-spin states. In the mix spin (MS) configuration, one of the cobalt is set to be in the high-spin state and the other three cobalt atoms are in the low-spin state.

we show the atomic projected density of states (PDOS) in Fig S5. Panel (a) is the PDOS of the LaCoO$_3$ thin film under a LaAlO$_3$ substrate (compressive strain). The calculations are done in a low-spin state, which yields a gap of 1.63 eV. The conduction band edge is dominantly Co-$d$ states, while the valence band edge is mainly O-$p$ states that are strongly hybridized with Co-$d$ states. Panel (b) is the PDOS of the LaCoO$_3$ thin film under a SrTiO$_3$ substrate (tensile strain). Motivated by the experimental results, we calculate a mixed-spin state in which 75% Co atoms are in a low-spin state and the 25% Co atoms are in a high-spin state. We find that the mixed-spin state is also insulating with a slightly smaller gap of 1.48 eV. The high-spin Co-$d$ state (red curve) has a strong splitting between spin-up and spin-down channel and yields a magnetic moment of 3.2$\mu_B$/Co. The splitting is roughly on the order of $U_{Co}$ = 6 eV. The low-spin Co-$d$ state (green curve) does not generate a net magnetization. The mixed-state has a net magnetization of 1$\mu_B$ per formula.

**ACKNOWLEDGMENTS**

This work was funded by the National Natural Science Foundation of China (Grant No. U1932116), Guangdong Basic and Applied Basic Research Foundation (Grant No. 2020B1515020029), Shenzhen Science and Technology Innovation project (Grant No. JCYJ20200109112829287), and Shenzhen Science and Technology Program (Grant No. KQTD20200820113045083). Z.H.C. has been supported by "the Fundamental Research Funds for the Central Universities" (grant no HIT.OCEF.2022038). C.-Y.K. acknowledge the financial support from the Ministry of Science and Technology in Taiwan under grant nos. MOST 110-2636-M-006-003 (Young Scholar Fellowship Program, Einstein Program). C.F.C. and C.-Y.K. acknowledge support from the Max Planck-POSTECH-Hsinchu Center for Complex Phase Materials. Q.H.Z. acknowledge the financial support from National Natural Science Foundation of China (Grant Nos. 52072400 and 52025025) and Beijing Natural Science Foundation (Z190010). H.C. is supported by the Ministry of Science and Technology of China under project number 2021YFE0107900.

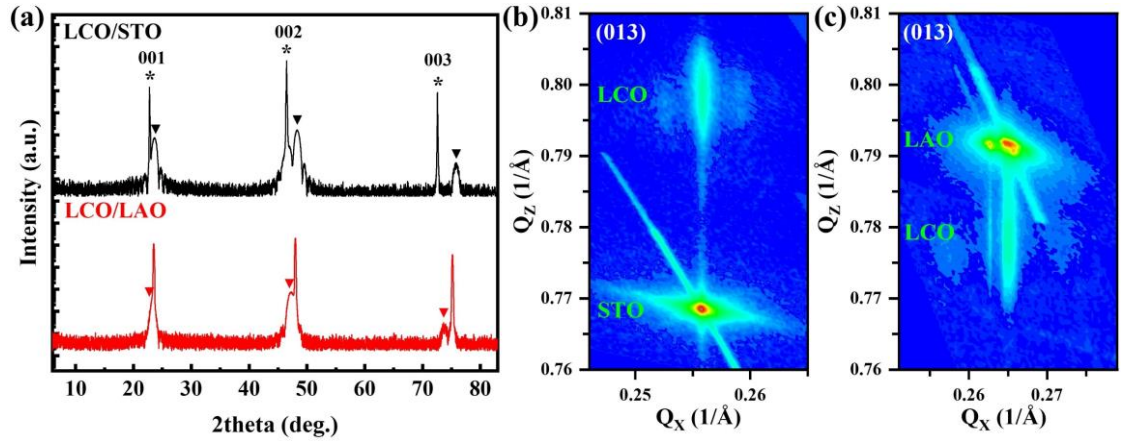

**FIG. 1.** Structural properties of LCO films grown on STO and LAO substrates. (a) X-ray diffraction $2\theta$-$\omega$ scan around the 001 and 002 reflection of LCO films grown on STO and LAO substrates. The peaks of substrates and LCO films are indicated with asterisk "*" and inverted triangle "▼", respectively. (b-c) Reciprocal space mappings (RSM) of LCO films grown on STO and LAO around the substrate's (013) reflection, respectively.

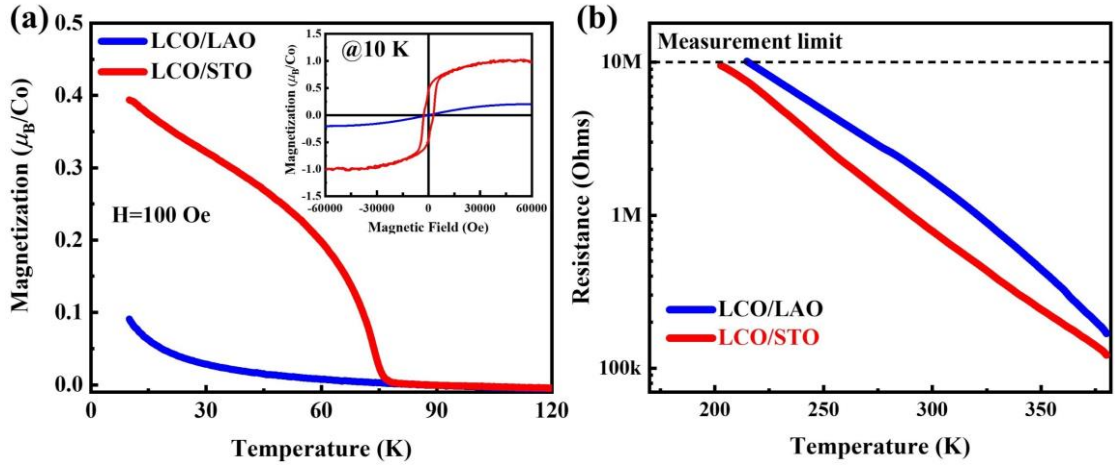

**FIG. 2.** (a) Magnetization-temperature (*M-T*) curves and magnetic hysteresis (*M-H*) loops measured at 10 K (the inset) of LCO films grown on LaAlO$_3$(001) and SrTiO$_3$(001) substrates. The *M-T* curves were measured at the magnetic field of 100 Oe applied along the film plane. (b) Temperature dependent resistance measured under zero magnetic field for 12 nm LaCoO$_3$ films grown on LaAlO$_3$(001) and SrTiO$_3$(001) substrate, respectively.

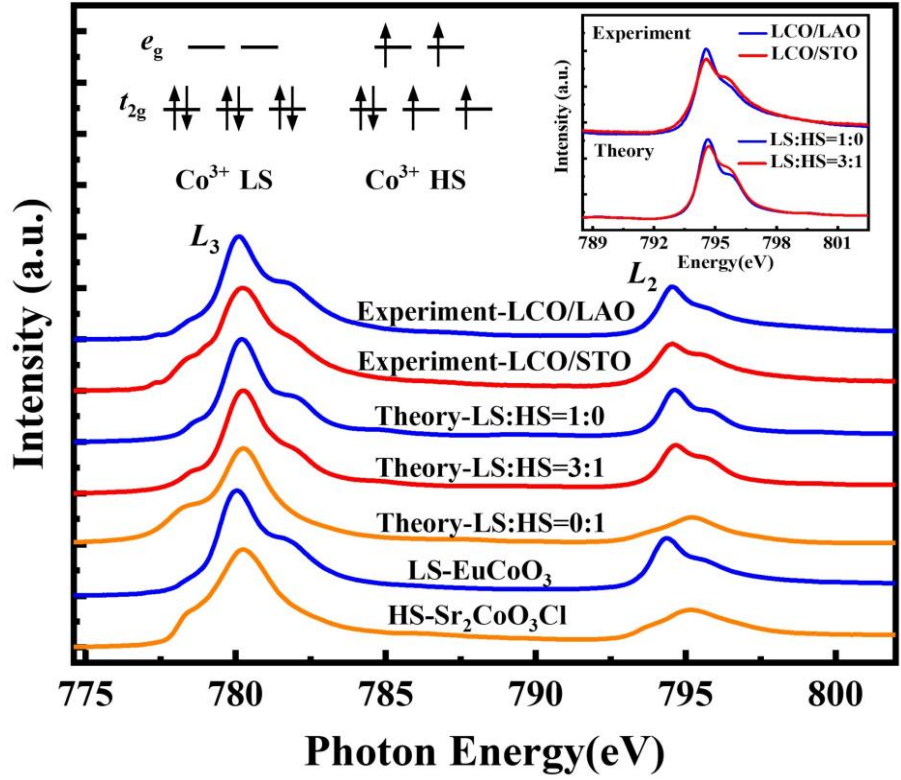

**FIG. 3.** Experimental Co $L_{2,3}$ XAS spectra taken from LCO films grown on LAO and STO substrates at 30 K, together with the theoretical XAS spectra in the LS-HS scenario and reference spectra for LS $Co^{3+}$ ($EuCoO_3$) and HS $Co^{3+}$ ($Sr_2CoO_3Cl$). Inset: The experimental Co $L_2$ XAS spectra of LCO/LAO and LCO/STO at 30 K and the corresponding theoretical isotropic spectra in the LS-HS scenario.

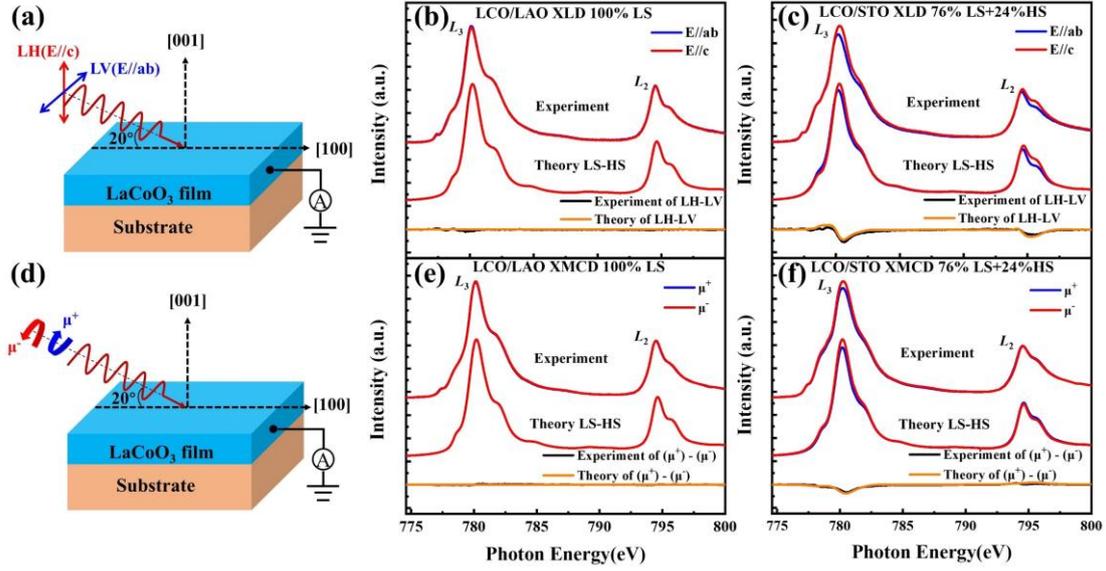

**FIG. 4.** Left panels: A sketch of the (a) XLD and (d) XMCD measurements. The XLD was obtained from the difference for XAS measured with linear horizontal and linear vertical polarized light. The XMCD was obtained from the difference for XAS measured with left ($\mu^+$) and right($\mu^-$) circularly polarized light. Before the XMCD measurement, the LCO films are first magnetized by a magnetic field of 1 Tesla. Both XLD and XMCD measurement, the X-ray beam was incident on the sample at an angle of 20° from the sample surface. Right panels: Top: Experimental Co $L_{2,3}$, XAS spectra and XLD at 30 K for (b) LCO/LAO and (c) LCO/STO, together with the corresponding theoretical XAS and XLD spectra calculated in the LS-HS scenario. Bottom: Experimental Co $L_{2,3}$ XAS spectra and XMCD at 30 K for (e) LCO/LAO and (f) LCO/STO using circularly polarized X-rays, together with the theoretical XAS and XMCD spectra for LCO/LAO and LCO/STO calculated in the LS-HS scenario.

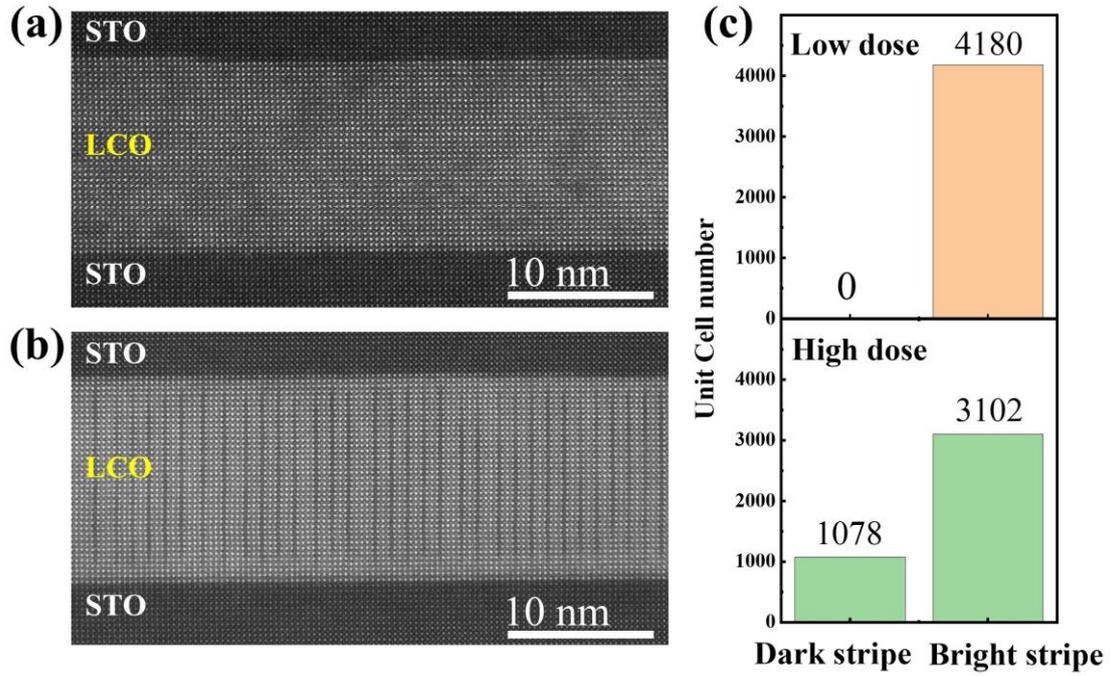

**FIG. 5.** High-resolution HAADF STEM image of a LCO film grown on a STO substrate with a STO capping layer on top with different electron dose condition; (a) low- and (b) high-dose conditions. (c) Unit cell number containing dark and bright stripe at low and high dose condition, respectively.

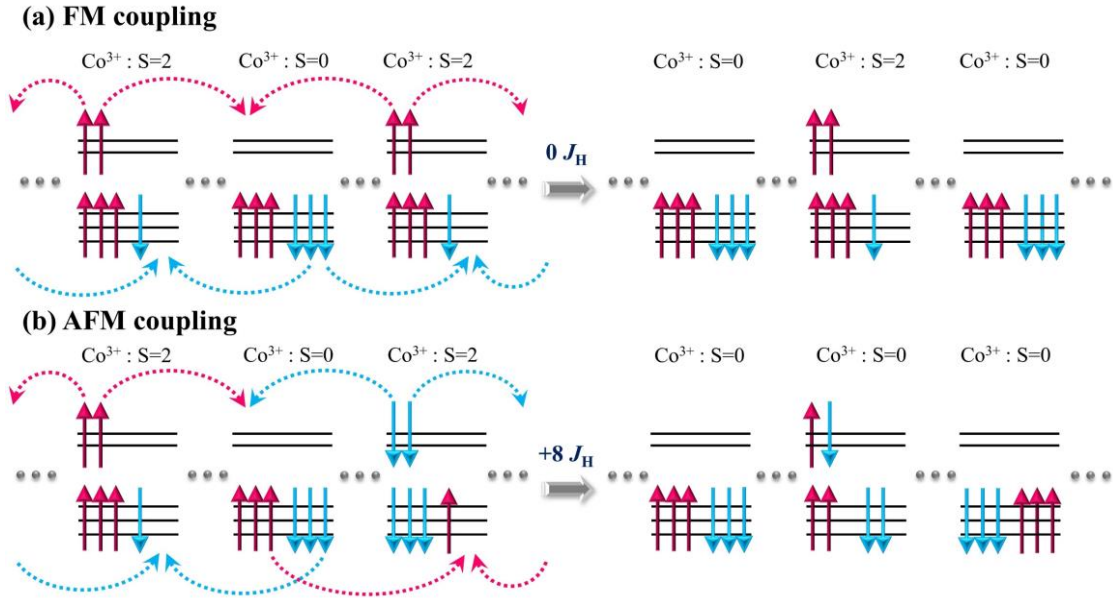

**FIG. 6.** The schematic diagram of the magnetic interactions between HS and LS $Co^{3+}$ ions with (a) ferromagnetic and (b) antiferromagnetic coupling. Its virtual hopping processes are as follows. The two $e_g$ electrons of a HS $Co^{3+}$ site hop evenly left and right to the neighboring LS $Co^{3+}$ sites. Two of the six $t_{2g}$ electrons of each LS $Co^{3+}$ site do a similar hopping but among $t_{2g}$ orbitals. This virtual hopping process prefers a ferromagnetic coupling because an antiferromagnetic coupling will lead to a wrong spin state configuration with an additional energy cost of $8J_H$ in respect to a ferromagnetic coupling. The blue dots represent LS $Co^{3+}$ sites. The number of the LS $Co^{3+}$ intermediate between the HS $Co^{3+}$ sites can be altered.